\documentclass[11pt]{article}
\usepackage{graphicx}
\usepackage{amssymb}

\begin{document}



\title{Storage Capacity Diverges with Synaptic Efficiency
in an Associative Memory Model
with Synaptic Delay and Pruning}

\author{
Seiji Miyoshi\thanks{
S.~Miyoshi is with the Department of Electronic Engineering, 
Kobe City College of Technology, Kobe 651-2194, Japan
(e-mail: miyoshi@kobe-kosen.ac.jp).}
, {\it Member, IEEE, }
and Masato Okada
\thanks{M.~Okada is with the
Laboratory for Mathematical Neuroscience,
RIKEN Brain Science Institute,
2-1, Hirosawa, Wako, Saitama 351-0198, Japan,
with ERATO Kawato Dynamic Brain Project, 
Japan Science and Technology Corporation, 
Kyoto 619-0288, Japan, and with 
Intelligent Cooperation and Control, PRESTO, 
Japan Science and Technology Corporation
(e-mail: okada@brain.riken.go.jp).}
}
\date{}


\maketitle

\centerline{Abstract}
It is known that storage capacity per synapse
increases by synaptic pruning
in the case of a correlation-type associative memory model.
However, the storage capacity of the entire network then decreases.
To overcome this difficulty, we propose decreasing the
connecting rate while keeping the total number of synapses 
constant by introducing delayed synapses.
In this paper, a discrete synchronous-type
model with both delayed synapses and their prunings 
is discussed as a concrete example of the proposal.
First, we explain the Yanai-Kim theory 
by employing the statistical neurodynamics.
This theory involves 
macrodynamical equations for the
dynamics of a network with serial delay elements.
Next, considering the translational symmetry of 
the explained equations, we re-derive
macroscopic steady state equations of the model 
by using the discrete Fourier transformation.
The storage capacities are analyzed quantitatively.
Furthermore, two types of synaptic prunings are
treated analytically: 
random pruning and systematic pruning.
As a result, it becomes clear that in both prunings, 
the storage capacity increases 
as the length of delay increases and 
the connecting rate of the synapses decreases 
when the total number of synapses is constant. 
Moreover, 
an interesting fact becomes clear:
the storage capacity asymptotically approaches $2/\pi$
due to random pruning. 
In contrast, the storage capacity diverges in proportion to 
the logarithm of the length of delay by systematic pruning
and the proportion constant is $4/\pi$.
These results theoretically support the significance of pruning 
following an overgrowth of synapses in the brain and 
strongly suggest that the brain prefers 
to store dynamic attractors such as sequences and limit cycles 
rather than equilibrium states.

\centerline{keywords}
associative memory, neural network, delay, synaptic pruning, 
statistical neurodynamics

\section{Introduction}
Robustness against noise and damage is often given 
as a positive feature of neural networks.
Therefore, it is important to analyze 
neural networks with respect to synaptic pruning.
Particularly in the case of correlation-type 
associative memory \cite{Okada1996},
models with randomly pruned synapses 
have been discussed in detail 
\cite{Sompo1986,Okada1998,OMK,Chechik1998}．
As a result, 
it became quantitatively clear
that synapse efficiency, which is 
defined by storage capacity per synapse, 
increases by synaptic pruning,
although storage capacity of the entire network
decreases.

On the other hand, it has often been observed
that synapses are pruned following an overgrowth
in real neural systems \cite{Chechik1998,Hutten1979,
Hutten1982,Bourgeois1993,Takacs1994,Innocenti1995,
Eckenhoff1991,Rakic1994,Stryker1986,Roe1990,Wolff1995}.
Though the functional significance of this phenomenon
is not known, Chechik et al. recently proposed the following hypothesis
\cite{Chechik1998}.
They considered cutting synapses that are lightly weighted
after learning with an excess of synapses,
expecting synapse efficiency to increase
by such systematic pruning.
Therefore, they hypothesized that increasing synapse
efficiency in this way adds functional significance
to synaptic pruning following an overgrowth.
They used a correlation-type auto-associative memory model
to verify this hypothesis.
After correlation learning,
they left heavily weighted synapses in the model, in which 
all neurons are fully connected.
Through computer simulations,
they showed that the synapse efficiency increased
by obtaining storage capacity.
In this paper, synaptic pruning as described above
is called systematic pruning.
Although the hypothesis of Chechik et al.~ is 
interesting as neuroscience, 
there are some unclear or imperfect points from the 
theoretical viewpoint: for example, what degree of
systematic pruning is more efficient than random pruning.
Accordingly, Mimura et al.~\cite{OMK} 
analyzed this system by using the self-consistent 
signal to noise analysis (SCSNA) \cite{Shiino1992}, 
which is a method of statistical mechanics. 
They showed that systematic pruning increased 
synapse efficiency by the order of $- \ln (1-R)$
over random pruning at the limit when $R$ approached unity, 
where $R \; (0 \geq R \geq 1)$
was the rate of synaptic pruning.
The important point in this case is that
the storage capacity of the entire network decreased,
though synapse efficiency increased by random pruning 
or systematic pruning.

To overcome this difficulty, we propose decreasing the
connecting rate while keeping the total number of synapses 
constant by introducing delayed synapses
with respect to a discrete synchronous-type model.
In this model, the storage capacity is expected to grow
with increases in synapse efficiency
because synapse efficiency increases by synaptic pruning, 
while the total number of synapses remains constant.
The discrete synchronous-type model with delayed synapses
\cite{Fukushima1973,YanaiNC,Miyoshi1995}
was proposed by Fukushima \cite{Fukushima1973}.
Yanai and Kim \cite{YanaiNC} theoretically analyzed this model 
with the statistical neurodynamics \cite{Amari1988}.
Their theory closely agrees with the results
of our computer simulation.

In this paper, after defining the model, 
we explain the Yanai-Kim theory \cite{YanaiNC,Miyoshi2002ICONIP,
Miyoshi2002NN} 
using the statistical neurodynamics \cite{Amari1988},
which involves macrodynamical calculations for a 
network with delayed synapses.
The Yanai-Kim theory needs a computational complexity of $O(L^4 t)$
to obtain the macrodynamics,
where $L$ and $t$ are the length of delay and the time step,
respectively.
Therefore, this theory is intractable
for discussing macroscopic properties
at the limit where $L$ is extremely large
\cite{Miyoshi2002ICONIP,Miyoshi2002}.
Thus, considering the translational symmetry
of time steps, which holds in the steady state
of the Yanai-Kim theory,
we re-derive the macroscopic steady state equations 
by employing the discrete Fourier transformation,
where the computational 
complexity does not formally depend on 
$L$ \cite{Miyoshi2002ICONIP,Miyoshi2002NN}.
Using the re-derived steady state equations, 
storage capacities can be quantitatively discussed
even for a large $L$ limit.

Next, synaptic pruning in the delayed network 
is investigated theoretically, and
storage capacities are evaluated quantitatively.
We deal with two types of pruning: random pruning and
systematic pruning.
As a result, it becomes clear that in both types of pruning, 
storage capacity increases as the length of delay increases,
while the connecting rate of synapses decreases
where the total number of 
synapses is constant. 
Moreover, 
an interesting fact becomes clear:
the storage capacity asymptotically approaches $2/\pi$
by random pruning. 
In contrast, the storage capacity diverges in proportion to 
the logarithm of the length of delay $L$,
that is, $(4/\pi)\ln L$,  by systematic pruning.

\section{Delayed Network}
\subsection{Model}
\begin{figure*}[htbp]
\begin{center}
\includegraphics[width=0.900\linewidth,keepaspectratio]{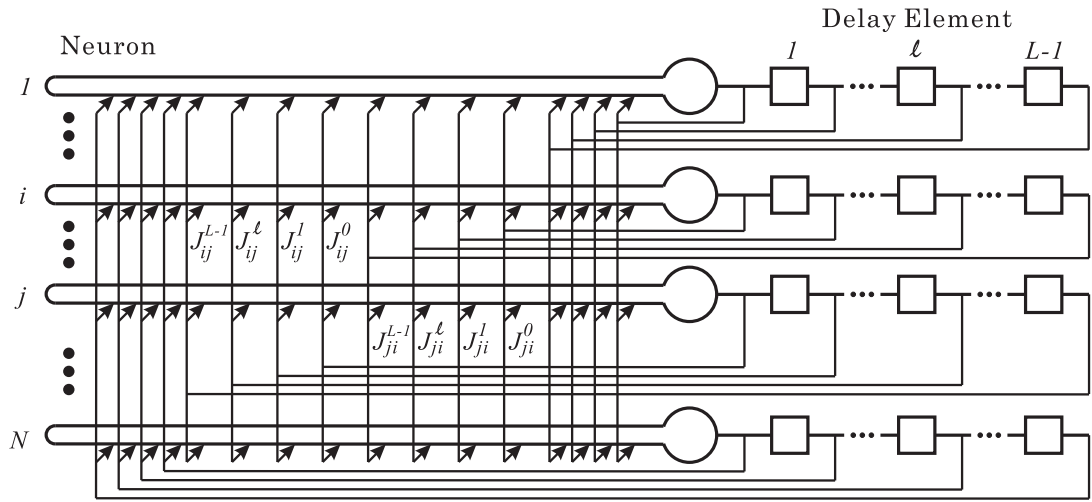}
\caption{Structure of delayed network.}
\label{fig:str}
\end{center}
\end{figure*}

The structure of the delayed network discussed in this paper is 
shown in Figure \ref{fig:str}. This figure corresponds to the case of 
fully synaptic connections, meaning no synaptic pruning.
The network has $N$ neurons, and $L-1$ serial delay elements are connected
to each neuron. All neurons, as well as all delay elements, have synaptic
connections with all neurons.
In this neural network, all neurons and all
delay elements change their states simultaneously,
i.e., this network employs a discrete synchronous updating rule.
The output of each neuron is determined by
\begin{eqnarray}
x_i^{t+1} &=& F\left(u_i^t\right) , \label{eqn:xF} \\
F\left(\cdot\right) &=& \mbox{sgn}\left(\cdot\right) , \\
u_i^t &=& \sum_{l=0}^{L-1}\sum_{j=1}^N J_{ij}^l x_j^{t-l},
\end{eqnarray}
where $x_i^t$ denotes 
the output of the $i$th neuron at time $t$, and $J_{ij}^l$
denotes the connection weight from the $l$th delay elements 
of the $j$th neuron to the $i$th neuron.
Here, $\mbox{sgn}$ is the sign function defined as
\begin{equation}
\mbox{sgn}\left(u\right)=\left\{
\begin{array}{ll}
+1,            & u \geq 0 ,  \\
-1,            & u <    0 .
\end{array}
\right.
\end{equation}
In this paper, the limit $N \rightarrow \infty$ is used
unless stated otherwise.

Let us consider the storing sequence of $\alpha N$
memory patterns, $\mbox{\boldmath $\xi$}^1 \rightarrow 
\mbox{\boldmath $\xi$}^2
\rightarrow \cdots \rightarrow \mbox{\boldmath $\xi$}^\mu
 \rightarrow \cdots \rightarrow 
\mbox{\boldmath $\xi$}^{\alpha N}$.
Here, $\alpha$ and $\alpha N$ are the loading rate
and the length of the sequence, respectively.
Each component of $\mbox{\boldmath $\xi$}^{\mu}$ is assumed to be an
independent random variable
that takes a value of either $+1$ or $-1$ according to
the following probabilities.
\begin{equation}
 \mbox{Prob}\left[\xi_i^{\mu}=\pm1\right]=\frac{1}{2} .
\end{equation}

The synaptic weight $J^l_{ij}$ is determined by correlation 
learning:
\begin{equation}
J_{ij}^l=\frac{c_l}{N}\sum_\mu \xi_i^{\mu+1+l}\xi_j^\mu,
\label{eqn:correlation}
\end{equation}
where $c_l$ is the strength of the $l$th delay step.

Correlation learning is an algorithm based 
on the Hebb rule, and it is inferior to the error correcting learning
in terms of storage capacity. However, as seen
in (\ref{eqn:correlation}), 
it is not necessary to 
re-learn all patterns that were
stored in the past when adding new patterns. 
Furthermore, correlation learning 
has been analyzed by many researchers due to its simplicity.

\subsection{Dynamical Behaviors of Macroscopic Order 
Parameters by Statistical Neurodynamics}
In the case of a small loading rate $\alpha$, 
if a state close to one or a set of the patterns 
stored as a sequence is given to the network,
the stored sequence of memory patterns is retrieved.
However, when the loading rate $\alpha$ increases, 
the memory fails at a certain $\alpha$. That is, 
even if a state close to one or a set of the patterns 
stored as a sequence is given to the network,
the state of the network tends to omit the stored 
sequence of memory patterns.
Moreover, even if one or a set of the patterns itself
is given to the network,
the state of the network tends to leave
the stored sequence of memory patterns.
This phenomenon of the memory suddenly
becoming unstable at a critical
loading rate can be considered a kind of phase transition.
Here, the storage capacity $\alpha_C$ is defined as 
the critical loading rate where recall becomes
unstable.

We define the overlap, or direction cosine, between 
a state $\mbox{\boldmath $x$}^t=\left(x_i^t\right)$ 
appearing in a recall process at time $t$ 
and an embedded pattern $\mbox{\boldmath $\xi$}^\mu
=\left(\xi_i^\mu\right)$ as
\begin{equation}
m_\mu^t=\frac{1}{N}\sum_{i=1}^N \xi_i^{\mu} x_i^t.
\label{eqn:mmut}
\end{equation}

Using this definition, when the state of the network 
at time $t$ and the $\mu$th pattern agree perfectly, 
the overlap $m_t^{\mu}$ is equal to unity.
When they have no
correlation, the overlap $m_t^{\mu}$ is equal to zero.
Therefore, the overlap provides a means of measuring recall quality.

Amari and Maginu
proposed the statistical neurodynamics
for the associative memory model \cite{Amari1988,Okada1995,Amari1988b}.
This analytical method handles the 
dynamical behavior of the associative memory model macroscopically,
where cross-talk noise is regarded as a Gaussian random
variable with a mean of zero and a time-dependent
variance of $\sigma_t^2$. They then derived recursive relations
for the variance and the overlap.

Yanai and Kim applied this method to the present model,
and succeeded in obtaining macroscopic state transition 
equations\cite{YanaiNC}. We will briefly explain their derivation
as follows.

The total input of the $i$th neuron at time $t$ is given as
\begin{eqnarray}
u_i^t
 &=& \sum_{l=0}^{L-1}\sum_{j=1}^N J_{ij}^lx_j^{t-l} \nonumber \\
 &=& s^t \xi_i^{t+1}+z_i^t , \label{eqn:sz} \\
s^t
 &=& \sum_{l=0}^{L-1} c_l m_{t-l}^{t-l} , \\
z_i^t
 &=& \sum_{l=0}^{L-1}c_l\sum_{\nu \neq t}
     \xi_i^{\nu+1}m_{\nu-l}^{t-l} . \label{eqn:zit}
\end{eqnarray}
The first term in (\ref{eqn:sz}) is the signal useful
for recall, while the second term is cross-talk noise that 
prevents $\xi_i^{t+1}$ from being recalled.
This procedure is called a signal-to-noise analysis.

We can then use the first-order Taylor expansion 
regarding $F(\cdot)$ to obtain
\begin{eqnarray}
m_\mu^t
&=& \frac{1}{N}\sum_{i=1}^N \xi_i^\mu x_i^t \nonumber \\
&=& \bar{m}_\mu^t + U_t\sum_{l'=0}^{L-1}c_{l'}m_{\mu-l'-1}^{t-l'-1} ,
\label{eqn:mmut2} \\
\bar{m}_\mu^t
&=& \frac{1}{N}\sum_{i=1}^N \xi_i^\mu F\left(\sum_{l=0}^{L-1}
    \sum_{j=1}^N\frac{c_l}{N}\right. \nonumber \\
& & \hspace{10mm} \times \left. 
    \sum_{\nu \neq \mu-l-1}\xi_i^{\nu+1+l}\xi_j^\nu x_j^{t-l-1}\right) ,\\
U_t
&=& \frac{1}{N}\sum_{i=1}^N 
    F'\left(\sum_{l=0}^{L-1}\sum_{j=1}^N\frac{c_l}{N}\right. 
    \nonumber \\
& & \hspace{10mm} \times \left.\sum_{\nu \neq \mu-l-1}\xi_i^{\nu+1+l}
    \xi_j^\nu x_j^{t-l-1}\right),
    \label{eqn:Ut}
\end{eqnarray}
where $F'(\cdot)$ is the differentiation of $F(\cdot)$.

Taking the correlation in the cross-talk noise
$z_i^t$ into account, we can derive the following macrodynamical 
equations using (\ref{eqn:xF})-(\ref{eqn:Ut})
(see Appendix A).
\begin{eqnarray}
\sigma_t^2
&=& \sum_{l=0}^{L-1}\sum_{l'=0}^{L-1}c_l c_{l'}v_{t-l,t-l'} ,
\label{eqn:sigma2} \\
v_{t-l,t-l'}
&=& \alpha \delta_{l,l'} \nonumber \\
&+& U_{t-l}U_{t-l'} \nonumber \\
& & \times 
    \sum_{k=0}^{L-1}\sum_{k'=0}^{L-1}c_k c_{k'}v_{t-l-k-1,t-l'-k'-1}
    \nonumber \\
&+& \alpha \left(c_{l-l'-1}U_{t-l'} + c_{l'-l-1}U_{t-l}\right) ,
\label{eqn:vD} \\
U_t &=& \sqrt{\frac{2}{\pi}} \frac{1}{\sigma_{t-1}}
        \exp\left(-\frac{\left(s^{t-1}\right)^2}{2\sigma_{t-1}^2}\right) ,
\label{eqn:U} \\
s^t&=&\sum_{l=0}^{L-1} c_l m_{t-l} , \\
m_{t+1}&=& \mbox{erf}\left(\frac{s^t}{\sqrt{2}\sigma_t}\right),
\label{eqn:mt1}
\end{eqnarray}
where $m_{t}$ denotes $m_{t}^{t}$.
$v_{t-l,t-l'}=\sum_{\mu\neq t}m_{\mu-l}^{t-l}m_{\mu-l'}^{t-l'}$.
$\sigma_t^2$ is the variance of the cross-talk noise.
$U_t$ is a kind of susceptibility, which measures the sensitivity of 
neuron output with respect to the external input.
If $t<0$, $m_t=0$ and $U_t=0$.
If $k<0$, $c_k=0$.
If either $k<0$ or $k'<0$, $v_{k,k'}=0$.
The expression
${\rm erf}\left(x\right)\equiv \frac{2}{\sqrt{\pi}}\int_0^x 
\exp\left(-u^2\right)du$
denotes the error function.
In this paper, the initial condition is that 
the states of all neurons and all delay elements are set to
be the stored pattern sequences.
In this case, 
$m_l=1 \ (l=0,\cdots,L-1)$ and 
$v_{l,l}=\alpha \ (l=0,\cdots,L-1)$.


\subsection{Macroscopic Steady State Analysis by Discrete
Fourier Transformation and Discussion}
The Yanai-Kim theory explained in the previous section,
which involves 
the macrodynamical equations obtained by the statistical 
neurodynamics, needs a computational complexity of
$O(L^4t)$ to obtain the macrodynamics
shown in (\ref{eqn:sigma2}) and (\ref{eqn:vD}),
where $L$ and $t$ are the length of
delay and the time step, respectively \cite{YanaiNC,
Miyoshi2002ICONIP,Miyoshi2002NN}.
Therefore, in this method,
it is difficult to investigate the critical loading rate 
for a large $L$ limit, i.e., the 
asymptotic behavior of the storage capacity in a 
large $L$ limit.
Thus, Miyoshi, Yanai and Okada considered the Yanai-Kim theory
in a steady state and derived the macroscopic steady
state equations of the delayed network. 
Furthermore, the storage capacity was analyzed
for a large $L$ by solving the derived equations 
numerically \cite{Miyoshi2002ICONIP,Miyoshi2002NN}.

We will briefly explain the derivation of the macroscopic
steady state equations to make the present paper self-contained.

For simplicity, let us assume that $c_l=1,\ l=0,\cdots,L-1$.
In a steady state, $v_{t-l,t-l'}$ can be expressed as $v_{l-l'}$
because of the translational symmetry in terms of time step.
Therefore, by modifying (\ref{eqn:sigma2}) and (\ref{eqn:vD}), 
we obtain
\begin{eqnarray}
\sigma^2 &=& \sum_{n=1-L}^{L-1}\left(L-|n|\right)v\left(n\right) , 
\label{eqn:sigmasteady}\\
v\left(n\right) &=& \alpha \delta_{n,0} \nonumber \\
     &+& U^2\sum_{i=1-L}^{L-1}\left(L-|i|\right)v\left(n-i\right)
         +U\alpha d\left(n\right) , \label{eqn:vsteady}\\
d\left(n\right) &=& \left\{
   \begin{array}{ll}
      1,            & |n|=1,2,\cdots,L  \\
      0,            & \mbox{otherwise}
   \end{array}
   \right.
\label{eqn:dn}
\end{eqnarray}
where $n=l-l' , i=k-k'$, $v(n)$ denotes $v_{n}$
and $\delta$ is Kronecker's delta.

Using the discrete Fourier transformation, 
%
%
we can obtain the steady state equations 
in terms of the network's macroscopic variables
as (\ref{eqn:sigma2steady})-(\ref{eqn:mm})
(see Appendix B).
\begin{figure*}[hbt]
\begin{equation}
\sigma^2
= \int_{-\frac{1}{2}}^{\frac{1}{2}}
\frac{\alpha \left[\left(1-U\right)\sin (\pi x) + 
U \sin \left\{\left(2L+1\right)\pi x\right\}\right]
\left[1-\cos (2L\pi x)\right]}
{\sin (\pi x) \left[2\sin^2(\pi x) 
-U^2\left\{1-\cos (2L\pi x)\right\}\right]}dx ,
\label{eqn:sigma2steady}
\end{equation}
\end{figure*}
\begin{eqnarray}
U &=& \sqrt{\frac{2}{\pi}}\frac{1}{\sigma}
    \exp\left(-\frac{s^2}{2\sigma^2}\right) ,
    \label{eqn:mUt} \\
s &=& mL , \label{eqn:s}\\
m &=& \mbox{erf}\left(\frac{s}{\sqrt{2}\sigma}\right) .
\label{eqn:mm}
\end{eqnarray}

Though the derived macroscopic steady state equations 
include a simple integral, their computational 
complexity does not formally depend on $L$.
Therefore, we can easily perform numerical calculations
for a large $L$.
Figure \ref{fig:Lsnocut2} shows the results of theoretical 
calculations in cases where $L=1,3$ and $10$, which are obtained 
by solving these equations numerically.
Figure \ref{fig:LdAs} shows the results of computer simulations.
In these Figures, the abscissa is the loading rate $\alpha$.
In the computer simulations, the number of neurons is $N=500$.
The initial condition is that 
the states of all neurons and all delay elements are set to
be the stored pattern sequences.
The steady state overlaps $m_{\infty}$ are 
obtained by calculations with a sufficient number of steps. 
Eleven simulations were carried out for each combination
of loading rate $\alpha$ and lengths of delay $L$.
Data points $\bullet$ , $\circ$ , {\tiny $\blacksquare$}
indicate the medians 
of the sixth largest values
for $L=$1,3 and 10, respectively, in the eleven trials. 
Error bars indicate the third and ninth 
largest values in the eleven trials.
In each trial, the loading rate is increased by adding 
new patterns.

These figures show that the steady states
obtained by the derived theory
agree closely with those obtained by computer simulation.
Therefore, in the case of a large $L$, 
only the theoretical calculations are executed.
Figure \ref{fig:Lsnocut} shows the results, while 
Figure \ref{fig:nocut} shows the relationship between
the length of delay $L$ and the storage capacity $\alpha_C$.

From these figures, we can see that the storage capacity 
increases in proportion to the length of delay $L$ with a 
large $L$ limit and a proportion constant of 0.195.
In other words, the storage capacity of the delayed network
$\alpha_C$ equals $0.195L$ when the length of delay $L$
is large \cite{Miyoshi2002ICONIP,Miyoshi2002NN}.
Although the result indicating that the delayed network's 
storage capacity
is in proportion to the length of delay $L$ may be trivial,
the fact that this result has been proven analytically is
significant. Moreover, the proportion constant 0.195 is 
a mathematically significant
number because 
it represents the limit of the delayed network's storage capacity.

\section{Synaptic Pruning}
\subsection{Necessity of Analyzing Synaptic Pruning} 
\label{sec:needs}
During brain development, 
the phenomenon of synaptic pruning following
overgrowth \cite{Chechik1998,Hutten1979,
Hutten1982,Bourgeois1993,Takacs1994,Innocenti1995,
Eckenhoff1991,Rakic1994,Stryker1986,Roe1990,Wolff1995}
can be observed.
Since this pruning following overgrowth seems to be
a universal phenomenon occurring in almost all areas
-- visual cortex, motor area, association area, and so on --
it is important to analyze synaptic pruning
and to discuss its properties quantitatively.

In real neural systems, 
some synaptic delay is inevitable. 
This property can be analyzed with a model 
that involves both delay elements and synaptic pruning.
For example, Figure \ref{fig:L5} shows that a
delay of three time steps can be represented by
pruning the first, second, fourth and fifth synapses,
and a five-time-step delay can be represented by
pruning the first, second, third and fourth synapses
with a model whose length of delay is five.
From this perspective, analyzing 
a model with both delay elements and synaptic pruning
is significant.

Moreover, in the case of a delayed network with no pruning,
it is obvious that storage capacity increases 
as the length of delay $L$ increases. 
On the contrary, it is interesting to analyze the
storage capacity of a delayed network that has a constant number of
synapses by introducing synaptic pruning.

It has been reported that the synapse efficiency, which is 
defined as storage capacity per synapse, 
increases due to synaptic pruning in networks with no delay
elements \cite{Sompo1986,Chechik1998}.
Two types of pruning can be considered, namely
random pruning and systematic pruning, which are typical methods
\cite{Sompo1986,Okada1998,Chechik1998}.
Mimura et al.~\cite{OMK} showed that 
synapse efficiency converged to $\frac{2}{\pi}$ by random pruning
and diverged as $\frac{2}{\pi}(-2\ln c)$ by systematic pruning
at the limit where
the connecting rate $c$ is extremely small.
Here, the relation between connecting rate $c$ and pruning rate $R$
was given by $c=1-R$.
The important point here is that
the storage capacity of the entire network decreases,
since the number of synapses decreases.

In the following discussion, a delayed network with 
synaptic pruning is analyzed on the basis of the macrodynamical
equations and macroscopic steady state equations re-derived
in the former section. 
We consider two types of pruning -- random 
pruning and systematic pruning -- for synaptic pruning.

\subsection{Random Pruning}
In this section, 
synapses of a delayed network are randomly pruned.
Random pruning of synapses can be realized without
any complicated control mechanism, so it is important
to investigate its effect on the dynamical behavior of 
pattern recall and storage capacity.

In the random synaptic pruning model, 
synaptic connections are constituted as
\begin{eqnarray}
J_{ij}^l
&=& \frac{c_lc_{ij}^l}{Nc}\sum_\mu \xi_i^{\mu+1+l}\xi_j^\mu,
\label{eqn:Jijl}
\end{eqnarray}
\begin{equation}
\mbox{Prob}[c_{ij}^l=1]= 1-\mbox{Prob}[c_{ij}^l=0]=c,
\end{equation}
where $c$ is the connecting rate.

Modifying (\ref{eqn:Jijl}), we obtain
\begin{eqnarray}
J_{ij}^l
&=& \frac{c_l}{N}\sum_\mu \xi_i^{\mu+1+l}\xi_j^\mu \nonumber \\
& & +\frac{c_l\left(c_{ij}^l-c\right)}{Nc}\sum_\mu \xi_i^{\mu+1+l}\xi_j^\mu .
\label{eqn:Jijl2}
\end{eqnarray}

Using (\ref{eqn:Jijl2}), we obtain the total input of 
the $i$th neuron at time $t$ as
\begin{eqnarray}
u_i^t
 &=& \sum_{l=0}^{L-1}\sum_{j=1}^N J_{ij}^lx_j^{t-l} \nonumber \\
 &=& s^t \xi_i^{t+1}+z_i^t \nonumber \\
 &+& \sum_{l=0}^{L-1}\sum_{j\neq i}\frac{c_l\left(c_{ij}^l-c\right)}{Nc}
     \sum_\mu \xi_i^{\mu+1+l}\xi_j^{\mu}x_j^{t-l} ,
 \label{eqn:random_u} \\
s^t
 &=& \sum_{l=0}^{L-1} c_l m_{t-l} ,\\
z_i^t
 &=& \sum_{l=0}^{L-1}c_l\sum_{\nu \neq t}
     \xi_i^{\nu+1}m_{\nu-l}^{t-l} .
\end{eqnarray}

As in the case of a fully connected network,
the first, second and third terms 
of (\ref{eqn:random_u})
are useful signals for recall, 
cross-talk noise 
and new noise generated by synaptic pruning, respectively.

In the third term of (\ref{eqn:random_u}),
\begin{eqnarray}
x_j^{t-l}
&=&F\left(u_j^{t-l-1}\right) \nonumber \\
&=&\sum_{l'=0}^{L-1}\sum_{k\neq j}J_{jk}^{l'}x_k^{t-l-1-l'} \nonumber \\
&=&F\left(\sum_{l'=0}^{L-1}\sum_{k\neq j}\frac{c_{l'}c_{jk}^{l'}}{Nc}
 \sum_{\nu}\xi_j^{\nu+1+l'}\xi_k^{\nu}x_k^{t-l-1-l'}\right) \nonumber \\
&=&x_j^{t-l,(\mu)}
+\left(\sum_{l'=0}^{L-1}\sum_{k\neq j}\frac{c_{l'}c_{jk}^{l'}}{Nc}
 \xi_j^{\mu+1+l'}\xi_k^{\mu}x_k^{t-l-1-l'}\right){x'}_j^{t-l,(\mu)},
\label{eqn:xjt-l}
\end{eqnarray}
where
\begin{eqnarray}
x_j^{t-l,(\mu)}
&=&F\left(\sum_{l'=0}^{L-1}\sum_{k\neq j}\frac{c_{l'}c_{jk}^{l'}}{Nc}
 \sum_{\nu\neq \mu}\xi_j^{\nu+1+l'}\xi_k^{\nu}x_k^{t-l-1-l'}\right) , \\
{x'}_j^{t-l,(\mu)}
&=&F'\left(\sum_{l'=0}^{L-1}\sum_{k\neq j}\frac{c_{l'}c_{jk}^{l'}}{Nc}
 \sum_{\nu\neq \mu}\xi_j^{\nu+1+l'}\xi_k^{\nu}x_k^{t-l-1-l'}\right) .
\end{eqnarray}
Using (\ref{eqn:xjt-l}), 
the third term of (\ref{eqn:random_u})
becomes
\begin{eqnarray}
& & \sum_{l=0}^{L-1}\sum_{j\neq i}\frac{c_l\left(c_{ij}^l-c\right)}{Nc}
    \sum_\mu \xi_i^{\mu+1+l}\xi_j^{\mu}x_j^{t-l} \nonumber \\
&=& \sum_{l=0}^{L-1}\sum_{j\neq i}\frac{c_l\left(c_{ij}^l-c\right)}{Nc}
    \sum_\mu \xi_i^{\mu+1+l}\xi_j^{\mu}x_j^{t-l,(\mu)} \nonumber \\
&+& \sum_{l=0}^{L-1}\sum_{j\neq i}\frac{c_l\left(c_{ij}^l-c\right)}{Nc}
    \sum_\mu \xi_i^{\mu+1+l}\xi_j^{\mu} \nonumber \\
& & \ \ \ \ \ \ \ \ \ \ \ \ \ \ \ \ \ 
    \sum_{l'=0}^{L-1}\sum_{k\neq j}\frac{c_{l'}c_{jk}^{l'}}{Nc}
    \xi_j^{\nu+1+l'}\xi_k^{\mu}x_k^{t-l-1-l'}{x'}_j^{t-l,(\mu)} .
\label{eqn:3term}
\end{eqnarray}

The second term of (\ref{eqn:3term}) becomes
\begin{equation}
\frac{1}{c^2}\sum_{l=0}^{L-1}\sum_{l'=0}^{L-1}
c_lc_{l'}\sum_{\mu}\xi_i^{\mu+1+l}
E\left[\xi_k^{\mu}x_k^{t-l-1-l'}\right]
E\left[\left(c_{ij}^l-c\right)c_{jk}^{l'}\xi_j^{\mu}{x'}_j^{t-l,(\mu)}\right],
\end{equation}
where $E\left[\xi_k^{\mu}x_k^{t-l-1-l'}\right]$ and
$E\left[\left(c_{ij}^l-c\right)c_{jk}^{l'}\xi_j^{\mu}{x'}_j^{t-l,(\mu)}\right]$
obey $N\left(0,1/N\right)$ and
$N\left(0,O\left(1/N\right)\right)$, respectively.
Therefore, the second term of (\ref{eqn:3term}) becomes $0$ for
a large $N$ limit. 
Here, $E[\cdot]$ and $N\left(a,\sigma^2\right)$
stand for an average and 
a Gaussian distribution with average $a$ and variance
$\sigma^2$, respectively.

Using this result and (\ref{eqn:3term}), the third term
of (\ref{eqn:random_u}) becomes \cite{Okada1998}
\begin{equation}
\sum_{l=0}^{L-1}\sum_{j\neq i}\frac{c_l\left(c_{ij}^l-c\right)}{Nc}
    \sum_\mu \xi_i^{\mu+1+l}\xi_j^{\mu}x_j^{t-l,(\mu)}
\sim N\left(0,\frac{\alpha(1-c)}{c}\sum_{l=0}^{L-1}c_l^2\right) .
\end{equation}

As a result, we can obtain the macrodynamical equations
for random pruning as follows.
\begin{eqnarray}
\tilde{\sigma}_t^2
&=& \sigma_t^2
    +\frac{\alpha \left(1-c\right)}{c}\sum_{l=0}^{L-1}c_l^2 ,
    \label{eqn:rsigma2}\\
\sigma_t^2
&=& \sum_{l=0}^{L-1}\sum_{l'=0}^{L-1}c_l c_{l'}v_{t-l,t-l'} ,\\
v_{t-l,t-l'}
&=& \alpha \delta_{l,l'} \nonumber \\
&+& U_{t-l}U_{t-l'} \nonumber \\
& & \times 
    \sum_{k=0}^{L-1}\sum_{k'=0}^{L-1}c_k c_{k'}v_{t-l-k-1,t-l'-k'-1}
    \nonumber \\
&+& \alpha \left(c_{l-l'-1}U_{t-l'} + c_{l'-l-1}U_{t-l}\right) ,\\
U_t &=& \sqrt{\frac{2}{\pi}} \frac{1}{\tilde{\sigma}_{t-1}}
  \exp\left(-\frac{\left(s^{t-1}\right)^2}{2\tilde{\sigma}_{t-1}^2}\right) , \\
s^t&=&\sum_{l=0}^{L-1} c_l m_{t-l} ,\label{eqn:st}\\
m_{t+1}&=&\mbox{erf}\left(\frac{s^t}{\sqrt{2}\tilde{\sigma}_t}\right),
\label{eqn:rm}
\end{eqnarray}
where the initial conditions are the same as in the case of 
a fully connected network. $\delta$ is Kronecker's delta.
Equation (\ref{eqn:rsigma2}) means that 
the variance $\tilde{\sigma}_t^2$ after pruning is 
the sum of the variance of cross-talk noise among patterns
and the variance of new noise generated by pruning.
Using the discrete Fourier transformation, 
as for a fully connected network, 
the macroscopic steady state equations in the case of
random pruning become
\begin{eqnarray}
\tilde{\sigma}^2
&=& \sigma^2
    +\frac{\alpha \left(1-c\right)}{c}\sum_{l=0}^{L-1}c_l^2 ,
    \label{eqn:rsigma2r} \\
U &=& \sqrt{\frac{2}{\pi}}\frac{1}{\tilde{\sigma}}
    \exp\left(-\frac{s^2}{2\tilde{\sigma}^2}\right) ,
    \label{eqn:mUt2} \\
s &=& mL , \label{eqn:s2}\\
m &=& \mbox{erf}\left(\frac{s}{\sqrt{2}\tilde{\sigma}}\right),
\label{eqn:mm2}
\end{eqnarray}
where $\sigma^2$ is given by (\ref{eqn:sigma2steady}).

It is obvious that storage capacity increases
with the length of delay $L$ 
if the connecting rate $c$ is constant.
Therefore, the storage capacity $\alpha_C$
is investigated under the condition 
that $c \times L$ is constant. 
This means that (\ref{eqn:rsigma2r})-(\ref{eqn:mm2})
are solved numerically and that the steady state overlaps $m_{\infty}$ 
are investigated by using $c=1/L$, where $c_l=1$.
Figure \ref{fig:LArt+s} shows the results of 
theoretical calculations and computer simulations 
when $L=1,2,3,5$ and $10$.
In this figure, the abscissa is loading rate $\alpha$.
In the computer simulations, the number of neurons is $N=500$.
The initial condition is that 
the states of all neurons and all delay elements are set to
be the stored pattern sequences.
The steady state overlaps $m_{\infty}$ are 
obtained by calculations with a sufficient number of steps. 
Eleven simulations were carried out for each combination
of loading rates $\alpha$ and lengths of delay $L$.
Data points $\bullet$ , $\circ$ , 
{\tiny $\blacksquare$} , {\tiny $\square$} , $*$
indicate the medians 
of the sixth largest values
for $L=$1,2,3,5 and 10, respectively,
in the eleven trials. 
Error bars indicate the third and the ninth 
largest values in the eleven trials.
In each subsequent trial, the loading rate is increased by adding 
new patterns.

Figure \ref{fig:LArt+s} displays the following results.
In the case of $L=1 \left(c=\frac{1}{L}=1.0\right)$,
which is fully connected with no delay elements,
the recurrent neural network's storage capacity $\alpha_C$ 
for sequential association is 0.269.
This agrees with the results of the previous works
\cite{Kawamura2002,During1998,Amari1988b}.
As the length of delay $L$ increases, storage capacity $\alpha_C$
increases
even though the total number of synapses is constant.
This phenomenon is due to the time lag 
of synaptic inputs by delays, which reduces
the statistical correlation among synaptic inputs.
As a result, variance of the noise component decreases.
This figure shows that theoretical results closely agree 
with the simulation results.
Therefore, only a theoretical calculation is executed 
when the length of delay $L$ is large, 
and the results of this calculation are shown in
Figure \ref{fig:randomt}.

The properties of $L=\infty$ in Figure \ref{fig:randomt} are
obtained as follows. The first term of the r.h.s. of (\ref{eqn:rsigma2r})
can be written as $\sigma^2=\alpha r$ from (\ref{eqn:sigma2steady}).
Therefore, we first numerically investigated the dependence of 
$r$ on $L$.
Figure \ref{fig:L-term1r} shows the results. The straight line in 
this figure shows a first-order approximation, which is obtained 
by using a least squares method, of the relation between
$\log L$ and $\log r$ at the phase transition point.
We can see that $r$ is $O(L^{1.45})$ by reading the slope of the line.
On the other hand, considering $c=1/L$ and $c_l=1$, when $L$ is 
extremely large, the second term of 
the r.h.s. of (\ref{eqn:rsigma2r}) becomes $\alpha L^2$,
that is $O(L^2)$. 
Therefore, only the second term is effective in the r.h.s. of
(\ref{eqn:rsigma2r}) when $L$ is extremely large,
and (\ref{eqn:rsigma2r}) becomes
\begin{equation}
\tilde{\sigma}^2 \rightarrow \alpha L^2 .
\label{eqn:aL2}
\end{equation}

Based on these considerations, the properties of $L=\infty$
in Figure \ref{fig:randomt} are obtained by ignoring the 
first term in the r.h.s. of (\ref{eqn:rsigma2r}).
Figure \ref{fig:randomt} shows that the steady state overlap
asymptotically approaches that of $L=\infty$ obtained
above as $L$ becomes large.

Now, in the case of $L=\infty$, 
the storage capacity can be obtained analytically as follows.
Substituting (\ref{eqn:s2}) into (\ref{eqn:mm2}), 
we obtain
\begin{equation}
m = \mbox{erf}\left(\frac{mL}{\sqrt{2}\tilde{\sigma}}\right) .
\label{eqn:s2mm2}
\end{equation}

Equation (\ref{eqn:s2mm2}) has 
nontrivial solutions $m\neq 0$ within the
range where the slope of the r.h.s. at $m=0$
is greater than 1.
Here, the slope of the r.h.s. of (\ref{eqn:s2mm2})
regarding $m$ can be written as
\begin{equation}
\frac{d}{dm}\mbox{erf}\left(\frac{mL}{\sqrt{2}\tilde{\sigma}}\right)
=\frac{L}{\tilde{\sigma}}\sqrt{\frac{2}{\pi}}
\exp\left(-\frac{m^2L^2}{2\tilde{\sigma}^2}\right) .
\label{eqn:slope}
\end{equation}

Therefore, we can obtain the critical value of the noise 
$\tilde{\sigma}^2_c$ as
\begin{equation}
\tilde{\sigma}^2_c=\frac{2}{\pi}L^2 .
\label{eqn:sigma_c}
\end{equation}

From (\ref{eqn:aL2}) and (\ref{eqn:sigma_c}),
the storage capacity $\alpha_C$ of random pruning 
at the limit when $L$ approaches $\infty$ is obtained as
\begin{equation}
\alpha_C=\frac{2}{\pi}\simeq 0.637 .
\end{equation}

Figure \ref{fig:randomt} shows that
the storage capacity approaches this value asymptotically
as $L$ increases.

\subsection{Systematic Pruning}
Chechik et al.\cite{Chechik1998} discussed 
the functional significance of 
synaptic pruning following overgrowth
on the basis of a correlation-type associative memory model.
They pointed out that synapse efficiency, which is storage capacity
per synapse, increases by cutting synapses that are lightly weighted
after correlation learning.

\begin{figure}[htbp]
\begin{center}
\includegraphics[width=0.400\linewidth,keepaspectratio]{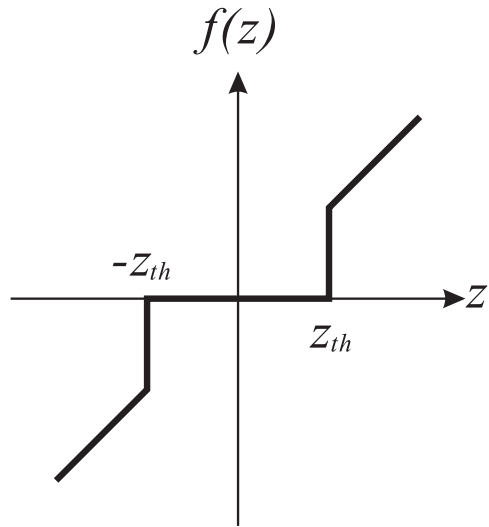}
\caption{Nonlinear function for systematic pruning.}
\label{fig:f}
\end{center}
\end{figure}

This type of systematic pruning can be expressed 
by nonlinear function $f(\cdot)$ shown in Figure \ref{fig:f}.
Synapses in the range of $-z_{th}<z<+z_{th}$ are pruned
by $f\left(\cdot\right)$.
In this case, 
synaptic connections are constituted by
\begin{eqnarray}
J_{ij}^l
&=& \frac{c_l\sqrt{\alpha N}}{N}f\left(T_{ij}^l\right) ,\\
T_{ij}^l
&=& \frac{1}{\sqrt{\alpha N}}\sum_\mu \xi_i^{\mu+1+l}\xi_j^\mu .
\label{eqn:Tij}
\end{eqnarray}
Equation (\ref{eqn:Tij}) is a stochastic variable that obeys 
normal distribution $N(0,1)$.
Therefore, the relationship between the connection rate $c$ and
$z_{th}$ is given by 
\begin{equation}
c=\int_{\{z|f\left(z\right)\neq 0\}}Dz
=1-\mbox{erf}\left(\frac{z_{th}}{\sqrt{2}}\right),
\label{eqn:c}
\end{equation}
where $Dz$ stands for 
$\frac{1}{\sqrt{2\pi}}\exp\left(-\frac{z^2}{2}\right)dz$, 
and the integral is from $-\infty$ to $+\infty$.

Modifying the connection weight $J_{ij}^l$\cite{Okada1998}, we obtain
\begin{eqnarray}
J_{ij}^l
&=& \frac{c_l\sqrt{\alpha N}}{N}f\left(T_{ij}^l\right) \nonumber \\
&=& \frac{c_l\sqrt{\alpha N}}{N}
\left(JT_{ij}^l+\left(f\left(T_{ij}^l\right)-JT_{ij}^l\right)\right),
\end{eqnarray}
where
\begin{eqnarray}
J
&=& \int Dx f'\left(x\right) \nonumber \\
&=& \int_{-\infty}^{\infty} \frac{dx}{\sqrt{2\pi}}\exp{-\frac{x^2}{2}}f'(x)
\nonumber \\
&=& \int_{-\infty}^{\infty} \frac{dx}{\sqrt{2\pi}}\exp{-\frac{x^2}{2}}xf(x)
\nonumber \\
&=& \int Dx xf\left(x\right).
\end{eqnarray}

Using these modification, we obtain the total input of 
the $i$th neuron at time $t$ is given as
\begin{eqnarray}
u_i^t
 &=& \sum_{l=0}^{L-1}\sum_{j=1}^N J_{ij}^lx_j^{t-l} \nonumber \\
 &=& \sum_{l=0}^{L-1}\sum_{j=1}^N \frac{c_l\sqrt{\alpha N}}{N}
   \left(JT_{ij}^l+\left(f\left(T_{ij}^l\right)-JT_{ij}^l\right)\right)
   x_j^{t-l} \\
 &=& \sum_{l=0}^{L-1}\sum_{j=1}^N \frac{c_lJ}{N}
   \sum_{\mu} \xi_i^{\mu+1+l}\xi_j^{\mu}x_j^{t-l} \nonumber \\
 &+& \sum_{l=0}^{L-1}\sum_{j=1}^N \frac{c_l \sqrt{\alpha N}}{N}
      \left(f\left(T_{ij}^l\right)-JT_{ij}^l\right)x_j^{t-l} \\
 &=& \left(J\sum_{l=0}^{L-1}c_lm_{t-l}^{t-l}\right)\xi_i^{t+1} \nonumber \\
 &+& J\sum_{l=0}^{L-1}c_l\sum_{\nu \neq t}
      \xi_i^{\nu+1}m_{\nu-l}^{t-l} \nonumber \\
 &+& \sum_{l=0}^{L-1}\sum_{j=1}^N \frac{c_l \sqrt{\alpha N}}{N}
      \left(f\left(T_{ij}^l\right)-JT_{ij}^l\right)x_j^{t-l} .
 \label{eqn:system_u}
\end{eqnarray}

As in the case of a fully connected network,
the first, second and third terms 
of (\ref{eqn:system_u})
are useful signals for recall, 
cross-talk noise 
and new noise generated by nonlinear transformation, respectively.
Here, the average of the third term equals $0$, and the variance 
equals \cite{Okada1998}
\begin{eqnarray}
& & E\left[\left(\sum_{l=0}^{L-1}\sum_{j=1}^N \frac{c_l 
     \sqrt{\alpha N}}{N}
     \left(f\left(T_{ij}^l\right)-JT_{ij}^l\right)x_j^{t-l}
     \right)^2\right] \nonumber \\
&=& E\left[\alpha\sum_{l=0}^{L-1}c_l^2\frac{1}{N}
    \sum_j\left(f\left(T_{ij}^l\right)-JT_{ij}^l\right)^2\right] \nonumber \\
&=& \alpha\sum_{l=0}^{L-1}c_l^2 
    \int Dx\left(f(x)-Jx\right)^2 \nonumber \\
&=& \alpha\sum_{l=0}^{L-1}c_l^2 
    \left(\int Dx f(x)^2 -2J\int Dxxf(x) + J^2 \int Dxx^2\right) \nonumber \\
&=& \alpha\sum_{l=0}^{L-1}c_l^2 
    \left(\int Dx f(x)^2 -2J^2 + J^2\right) \nonumber \\
&=& \alpha \left(\tilde{J}^2-J^2\right)\sum_{l=0}^{L-1}c_l^2,
\end{eqnarray}
where 
$\tilde{J}^2 = \int Dx\left(f\left(x\right)\right)^2$.
As a result, we can obtain the macrodynamical equations
for systematic pruning as follows.
\begin{eqnarray}
\tilde{\sigma}_t^2
&=& 
\sigma_t^2+\alpha \left(\tilde{J}^2-J^2\right)\sum_{l=0}^{L-1}c_l^2 ,
\label{eqn:ssigma2} \\
\sigma_t^2
&=& J^2 \sum_{l=0}^{L-1}\sum_{l'=0}^{L-1}c_l c_{l'}v_{t-l,t-l'} ,\\
\tilde{J}^2 &=& \int Dx\left(f\left(x\right)\right)^2 ,
\label{eqn:tildeJ2} \\
J &=& \int Dx xf\left(x\right) ,\label{eqn:J} 
\end{eqnarray}
\begin{eqnarray}
v_{t-l,t-l'}
&=& \alpha \delta_{l,l'} \nonumber \\
&+& U_{t-l}U_{t-l'} \nonumber \\
& & \times 
    \sum_{k=0}^{L-1}\sum_{k'=0}^{L-1}c_k c_{k'}v_{t-l-k-1,t-l'-k'-1}
    \nonumber \\
&+& \alpha \left(c_{l-l'-1}U_{t-l'} + c_{l'-l-1}U_{t-l}\right) ,
\end{eqnarray}
\begin{equation}
U_t = \sqrt{\frac{2}{\pi}}\frac{1}{\tilde{\sigma}_{t-1}}
    \exp\left(-\frac{\left(s^{t-1}\right)^2}{2\tilde{\sigma}_{t-1}^2}\right) ,
    \label{eqn:mUt3}
\end{equation}
\begin{equation}
s^t = J\sum_{l=0}^{L-1}c_lm_{t-l} ,
\label{eqn:stJ}
\end{equation}
\begin{equation}
m_{t+1} = \mbox{erf}\left(\frac{s^t}{\sqrt{2}\tilde{\sigma}_t}\right),
\label{eqn:mm3}
\end{equation}
where the initial conditions are the same as in the case of 
a fully connected network.  $\delta$ is Kronecker's delta.
Equation (\ref{eqn:ssigma2}) means that 
the variance $\tilde{\sigma}_t^2$ after pruning is 
the sum of the variance $\sigma_t^2$ of cross-talk noise among patterns
and the variance
$\alpha \left(\tilde{J}^2-J^2\right)\sum_{l=0}^{L-1}c_l^2$
of new noise generated by pruning.
Using the discrete Fourier transformation, 
as in the case of full connections, 
the macroscopic steady state equations in the case of
systematic pruning become
\begin{eqnarray}
\tilde{\sigma}^2
&=& 
\sigma^2+\alpha \left(\frac{\tilde{J}^2}{J^2}-1\right)\sum_{l=0}^{L-1}c_l^2 ,
    \label{eqn:rsigma2s} \\
U &=& \sqrt{\frac{2}{\pi}}\frac{1}{\tilde{\sigma}}
    \exp\left(-\frac{s^2}{2\tilde{\sigma}^2}\right) ,
    \label{eqn:mUt4} \\
s &=& mL ,\label{eqn:s3}\\
m &=& \mbox{erf}\left(\frac{s}{\sqrt{2}\tilde{\sigma}}\right),
\label{eqn:mm4}
\end{eqnarray}
where $\sigma^2$ is given by (\ref{eqn:sigma2steady}).

As for random pruning,
the storage capacity $\alpha_C$
is investigated under the condition 
that $c \times L$ is constant. 
This means that (\ref{eqn:rsigma2s})-(\ref{eqn:mm4})
are solved numerically, and the steady state overlaps $m_{\infty}$ 
are investigated by using $c=1/L$, where $c_l=1$.
Figure \ref{fig:LAst+s} shows the results of 
theoretical calculations and computer simulations
when $L=1,2,3,5$ and $10$.
In this figure, the abscissa is the loading rate $\alpha$.
In the computer simulations, the number of neurons is $N=500$,
and the steady state overlaps $m_{\infty}$ are 
obtained by calculations with a sufficient number of steps. 
Eleven simulations were carried out for each combination
of loading rate $\alpha$ and length of delays $L$.
Data points $\bullet$ , $\circ$ , 
{\tiny $\blacksquare$} , {\tiny $\square$} , $*$
indicate the medians 
of the sixth largest values
for $L=$1,2,3,5 and 10, respectively,
in the eleven trials. 
Error bars indicate the third and the ninth 
largest values in the eleven trials.
In each trial, the loading rate is increased by adding 
new patterns.

Figure \ref{fig:LAst+s} shows that
as the length of delay $L$ increases, storage capacity $\alpha_C$
increases, though the total number of synapses is constant.
This figure also shows that theoretical results closely agree 
with the simulation results.
Therefore, only a theoretical calculation is executed 
when the length of delay $L$ is large.
Figure \ref{fig:systemt} shows the results.
Figure \ref{fig:systemcapa} shows the relationship
between the length of delay $L$ and the storage capacities.

Here, we investigate the dependence of 
the first and second terms of 
the r.h.s. of (\ref{eqn:rsigma2s}) on $L$ 
in the same manner as random pruning
to obtain the asymptotic storage capacity analytically
when $L$ is extremely large. 

The first term of the r.h.s. of (\ref{eqn:rsigma2s})
can be written as $\sigma^2=\alpha r$ from (\ref{eqn:sigma2steady}).
Therefore, we first numerically investigate the dependence of 
$r$ on $L$.
Figure \ref{fig:L-term1s} shows the results. The straight line in 
this figure shows a first-order approximation, which is obtained 
by using a least squares method, of the relation between
$\log L$ and $\log r$ at the phase transition point.
We can see that $r$ is $O(L^{1.33})$ by reading the slope of the line.

On the other hand, the dependence of the second term on $L$
can be obtained as follows.
When $L$ is extremely large, that is, when $c=1/L$ is extremely
small and $z_{th}$ is extremely large, the connection rate $c$
of (\ref{eqn:c}) is as follows.
\begin{eqnarray}
c &=& \int_{\{z|f\left(z\right)\neq 0\}}Dz \nonumber \\
  &=& 1-\mbox{erf}\left(\frac{z_{th}}{\sqrt{2}}\right) \nonumber \\
  &\rightarrow& \sqrt{\frac{2}{\pi}}z_{th}^{-1}
                \exp\left(-\frac{z_{th}^2}{2}\right), 
                \ \ \ z_{th}\rightarrow \infty.
\label{eqn:casym}
\end{eqnarray}

$J$ and $\tilde{J}^2$ of (\ref{eqn:J}) and (\ref{eqn:tildeJ2})
become
\begin{eqnarray}
J &=& 2\int_t^{\infty}Dz z^2 \nonumber \\
  &=& \sqrt{\frac{2}{\pi}}z_{th}
      \exp\left(-\frac{z_{th}^2}{2}\right)
      +1-\mbox{erf}\left(\frac{z_{th}}{\sqrt{2}}\right) \nonumber \\
  &\rightarrow& \sqrt{\frac{2}{\pi}}z_{th}
                \exp\left(-\frac{z_{th}^2}{2}\right),
                \ \ \ z_{th}\rightarrow \infty, 
                \label{eqn:zthasym}
\\
\tilde{J}^2 
  &=& 2\int_t^{\infty}Dz z^2 \nonumber \\
  &=& J.\label{eqn:J2asym}
\end{eqnarray}

Considering $c=1/L$ and $c_l=1$, from (\ref{eqn:casym})，
(\ref{eqn:zthasym}) and (\ref{eqn:J2asym}),
the second term of (\ref{eqn:rsigma2s})
can be transformed as 
\begin{eqnarray}
\alpha
\left(\frac{\tilde{J}^2}{J^2}-1\right)\sum_{l=0}^{L-1}c_l^2
&=&\alpha
\left(\frac{1}{J}-1\right)L \nonumber \\
&\rightarrow& \alpha
         \left(\frac{1}{\sqrt{\frac{2}{\pi}}z_{th}
         \exp\left(-\frac{z_{th}^2}{2}\right)}-1\right)L \nonumber \\
&\rightarrow& \alpha \left(\frac{1}{z_{th}^{2}c}-1\right)L \nonumber \\
&\rightarrow& \alpha \frac{1}{z_{th}^2c}L \nonumber \\
&\rightarrow& \alpha \frac{1}{-2c\ln c}L \nonumber \\
&=& \alpha \frac{L^2}{2\ln L} . \label{eqn:aL2lnL}
\end{eqnarray}

Since (\ref{eqn:aL2lnL}) is $O(L^2)$,
only the second term is effective in the r.h.s. of 
(\ref{eqn:rsigma2s}) when $L$ is extremely large.
Based on these considerations, the storage capacity 
of systematic pruning can be obtained as follows.
Substituting (\ref{eqn:s3}) into (\ref{eqn:mm4}),
we obtain
\begin{equation}
m = \mbox{erf}\left(\frac{mL}{\sqrt{2}\tilde{\sigma}}\right) .
\label{eqn:s3mm4}
\end{equation}

Equation (\ref{eqn:s3mm4}) has 
nontrivial solutions $m\neq 0$ within the
range where the slope of the r.h.s. at $m=0$
is greater than 1.
Here, the slope of the r.h.s. of (\ref{eqn:s3mm4})
regarding $m$ can be written as
\begin{equation}
\frac{d}{dm}\mbox{erf}\left(\frac{mL}{\sqrt{2}\tilde{\sigma}}\right)
=\frac{L}{\tilde{\sigma}}\sqrt{\frac{2}{\pi}}
\exp\left(-\frac{m^2L^2}{2\tilde{\sigma}^2}\right) .
\label{eqn:slope2}
\end{equation}

Therefore, we can obtain the critical value of the noise 
$\tilde{\sigma}^2_c$ as
\begin{equation}
\tilde{\sigma}^2_c=\frac{2}{\pi}L^2 .
\label{eqn:sigma_c2}
\end{equation}

From (\ref{eqn:aL2lnL}) and (\ref{eqn:sigma_c2}),
the storage capacity $\alpha_C$ of systematic pruning 
at the limit when $L$ approaches $\infty$ is obtained as
\begin{equation}
\alpha_C=\frac{4}{\pi}\ln L .
\end{equation}

Figures \ref{fig:systemcapa} shows that 
the storage capacity of large $L$ is parallel
with the line of $\frac{4}{\pi}\ln L$ when $L$ is large.
This means that the storage capacity approaches $\frac{4}{\pi}\ln L$
relatively, and this result supports
the derived theory.

As the length of delay $L$ increases, storage capacity $\alpha_C$
increases, even though the total number of synapses remains constant;
the tendency of increase is different from that of random pruning.
Storage capacity is in proportion to the logarithm of the 
length of delay $L$, and the proportion constant is $\frac{4}{\pi}$.
In other words, for systematic pruning, storage capacity diverges 
with the increase in the length of delay $L$.
It is amazing that the storage capacity diverges regardless
of whether the total number of synapses is constant.

\section{Conclusions}
We analyzed a discrete synchronous-type model 
that adopts correlation learning by using 
the statistical neurodynamics and discussed
sequential associative memory by recurrent neural networks
with synaptic delay and pruning.
First, we explained the Yanai-Kim theory \cite{YanaiNC}, 
which involves macrodynamical equations for the
dynamics of a network with serial delay elements.
Next, considering the translational symmetry of 
the explained equations, 
we explained the macroscopic steady state equations of 
the model by using the discrete Fourier transformation
\cite{Miyoshi2002ICONIP,Miyoshi2002NN}.
The storage capacity was analyzed quantitatively.
As a result, we showed that the storage capacity is
in proportion to the length of delay $L$
when the $L$ limit is large and the proportion constant is 0.195.
Furthermore, two types of synaptic prunings were
analyzed: random pruning and systematic pruning.
As a result, it became clear that under both pruning conditions, 
the storage capacity grows with an increase in 
delay and a decrease in the connecting rate 
when the total number of synapses is constant. 
Moreover, 
an interesting fact became clear: the storage capacity 
approaches $2/\pi$ asymptotically by random pruning. 
In contrast, the storage capacity diverges in proportion to 
the logarithm of the length of delay by systematic pruning,
and the proportion constant is $4/\pi$.
These results theoretically support the significance of pruning 
following overgrowth of synapses in the brain
\cite{Chechik1998,Hutten1979,
Hutten1982,Bourgeois1993,Takacs1994,Innocenti1995,
Eckenhoff1991,Rakic1994,Stryker1986,Roe1990,Wolff1995} 
and strongly suggest that the brain prefers to store dynamic attractors 
such as sequences or limit cycles rather than equilibrium states.


\section*{Appendix A: Derivations of the Macrodynamical Equations of 
the Delayed Network} \label{sec:A1}
Using (\ref{eqn:zit}) and (\ref{eqn:mmut2}), we obtain
\begin{eqnarray}
z_i^t &=& z_A+z_B ,\label{eqn:zitAB} \\
z_A &=& \sum_{l=0}^{L-1}c_l\sum_{\mu\neq t} \xi_i^{\mu+1}\frac{1}{N}
\sum_j\xi_j^{\mu-l}x_j^{t-l,\left(\mu-l\right)} ,\label{eqn:zA} \\
z_B &=& \sum_{l=0}^{L-1}c_l\sum_{\mu\neq t} \xi_i^{\mu+1}
U_{t-l}\sum_{l'=0}^{L-1}c_{l'}m_{\mu-l-l'-1}^{t-l-l'-1},
\label{eqn:zB}
\end{eqnarray}
where $x_j^{t-l,\left(\mu-l\right)}$ is the variable 
obtained by removing
the influence of $\xi_j^{\mu-l}$ from $x_j^{t-l}$.
Using (\ref{eqn:zitAB}),(\ref{eqn:zA}) and (\ref{eqn:zB}),
we obtain
\begin{eqnarray}
E\left[z_i^t\right] &=& 0 ,\\
.\raisebox{1ex}{.}. \ \ \ \ \ 
\sigma_t^2 &=& E\left[\left(z_i^t\right)^2\right] \label{eqn:sigmat20} \\
 &=& E\left[z_A^2+z_B^2+2z_Az_B\right] .\label{eqn:sigmat2}
\end{eqnarray}

Transforming $z_A^2,z_B^2$ and $z_Az_B$ with consideration given to 
their correlation, we obtain
\begin{eqnarray}
E\left[z_A^2\right] &=& \alpha \sum_{l=0}^{L-1}c_l^2 \label{eqn:A2} ,\\
E\left[z_B^2\right] &=& \sum_{\mu\neq t}\sum_{l=0}^{L-1}\sum_{l'=0}^{L-1}
\sum_{k=0}^{L-1}\sum_{k'=0}^{L-1}c_lc_{l'}c_kc_{k'} \nonumber \\
 & & \times U_{t-l}U_{t-l'}m_{\mu-l-k-1}^{t-l-k-1}m_{\mu-l'-k'-1}^{t-l'-k'-1} ,
\label{eqn:B2} \\
E\left[2z_Az_B\right] &=& 
 \alpha\sum_{l=0}^{L-1}\sum_{l'=0}^{L-1}c_lc_{l'} \nonumber \\
& & \times \left(c_{l-l'-1}U_{t-l'}+c_{l'-l-1}U_{t-l}\right),
\label{eqn:2AB}
\end{eqnarray}
where
\begin{equation}
v_{t-l,t-l'}=\sum_{\mu\neq t}m_{\mu-l}^{t-l}m_{\mu-l'}^{t-l'} .
\end{equation}
Using (\ref{eqn:sigmat2})-(\ref{eqn:2AB}),
we obtain
\begin{eqnarray}
\sigma_t^2 &=& \alpha \sum_{l=0}^{L-1}c_l^2 \nonumber \\
&+& \sum_{l=0}^{L-1}\sum_{l'=0}^{L-1}
\sum_{k=0}^{L-1}\sum_{k'=0}^{L-1}c_lc_{l'}c_kc_{k'} \nonumber \\
 & & \times U_{t-l}U_{t-l'}v_{t-l-k-1,t-l'-k'-1}
\nonumber \\
&+& \alpha\sum_{l=0}^{L-1}\sum_{l'=0}^{L-1}c_lc_{l'}
\left(c_{l-l'-1}U_{t-l'}+c_{l'-l-1}U_{t-l}\right) .
\label{eqn:sigmat22}
\end{eqnarray}

Using (\ref{eqn:zit}) and (\ref{eqn:sigmat20}), we obtain
\begin{equation}
\sigma_t^2 = \sum_{l=0}^{L-1}\sum_{l'=0}^{L-1}c_lc_{l'}v_{t-l,t-l'} .
\label{eqn:sigmat23}
\end{equation}

Comparing (\ref{eqn:sigmat22}) and (\ref{eqn:sigmat23}) as
identical equations regarding $c_lc_{l'}$,
we obtain
\begin{eqnarray}
v_{t-l,t-l'}
&=& \alpha \delta_{l,l'} \nonumber \\
&+& U_{t-l}U_{t-l'} \nonumber \\
& & \times 
    \sum_{k=0}^{L-1}\sum_{k'=0}^{L-1}c_k c_{k'}v_{t-l-k-1,t-l'-k'-1}
    \nonumber \\
&+& \alpha \left(c_{l-l'-1}U_{t-l'} + c_{l'-l-1}U_{t-l}\right),
\label{eqn:v}
\end{eqnarray}
where $\delta$ is Kronecker's delta.
Using (\ref{eqn:Ut}), we obtain
\begin{eqnarray}
U_t
&=& \frac{1}{N}\sum_{i=1}^N 
    F'\left(\sum_{l=0}^{L-1}\sum_{j=1}^N\frac{c_l}{N}\right. 
    \nonumber \\
& & \hspace{10mm} \times \left.\sum_{\nu \neq \mu-l-1}\xi_i^{\nu+1+l}
    \xi_j^\nu x_j^{t-l-1}\right) \nonumber \\
&=& E\left[F'\left(u^{t,\left(\mu\right)}\right)\right] \nonumber \\
&=& E\left[F'\left(u^t\right)\right] \nonumber \\
&=& \int\frac{dz}{\sqrt{2\pi}}e^{-\frac{z^2}{2}}\ll F'\left(u^t\right)\gg
\nonumber \\
&=& \frac{1}{\sigma}\int\frac{dz}{\sqrt{2\pi}}e^{-\frac{z^2}{2}}z
\ll F\left(u^t\right)\gg \nonumber \\
&=& \sqrt{\frac{2}{\pi}} \frac{1}{\sigma_{t-1}}
    \exp\left(-\frac{\left(s^{t-1}\right)^2}{2\sigma_{t-1}^2}\right),
    \label{eqn:Ut2}
\end{eqnarray}
where $u^{t,\left(\mu\right)}$ is the variable 
obtained by removing the
influence of $\mbox{\boldmath $\xi$}^\mu$ from $u^{t}$.
Here, $\ll\cdot\gg$ stands for the average over pattern 
$\mbox{\boldmath $\xi$}$.

As a result, we can obtain the macrodynamical equations
for overlap $m$, that is,
(\ref{eqn:sigma2})-(\ref{eqn:mt1}).

\section*{Appendix B: Derivations of the Macroscopic Steady State
Equations by Discrete Fourier Transformation} \label{sec:A2}
Using the discrete Fourier transformation, we re-derive 
the general term of $v(n)$,
which is expressed
by the recurrence formula in (\ref{eqn:vsteady})
\cite{Miyoshi2002ICONIP,Miyoshi2002NN}.
Applying the discrete Fourier transformation to
(\ref{eqn:vsteady}) and (\ref{eqn:dn}), we obtain
\begin{eqnarray}
V(r) &=& \alpha+U^2\sum_{i=1-L}^{L-1}(L-|i|)V(r)
e^{-j2\pi \frac{ri}{2T+1}}+\alpha UD(r) ,\label{eqn:Vk}\\
D(r) &=& \sum_{n=-T}^T d(n)e^{-j2\pi \frac{rn}{2T+1}} \nonumber \\
&=& \sum_{n=1}^L 
\left(e^{-j2\pi \frac{rn}{2T+1}}+e^{j2\pi \frac{rn}{2T+1}}\right),
\label{eqn:Dk}
\end{eqnarray}
where $V(r)$ and $D(r)$ are the discrete Fourier transformations
of $v(n)$ and $d(n)$, respectively.

Solving (\ref{eqn:Vk}) and (\ref{eqn:Dk}) in terms of $V(r)$,
we obtain
\begin{equation}
V\left(r\right)=\frac{\alpha \left( 1+U\sum_{n=1}^L 
\left(e^{-j2\pi \frac{rn}{2T+1}}+e^{j2\pi \frac{rn}{2T+1}}\right)
\right)}
{1-U^2\sum_{i=1-L}^{L-1}(L-|i|)e^{-j2\pi \frac{ri}{2T+1}}} .
\label{eqn:Vk2}
\end{equation}

Summations in (\ref{eqn:Vk2}) are calculated as follows.
\begin{eqnarray}
& & \sum_{n=1}^L 
\left(e^{-j2\pi \frac{rn}{2T+1}}+e^{j2\pi \frac{rn}{2T+1}}\right)
\nonumber \\
&=& \left\{
 \begin{array}{ll}
  \frac{\sin \left((2L+1)\frac{\pi r}{2T+1}\right)}
       {\sin\left(\frac{\pi r}{2T+1}\right)}-1,    & r \ne 0 ,  \\
  2L,                                              & r=0 ,
 \end{array}
\right.
\label{eqn:Dr}
\end{eqnarray}

\begin{eqnarray}
\sum_{i=1-L}^{L-1}e^{-j2\pi \frac{ri}{2T+1}}
=\left\{
\begin{array}{ll}
  \frac{\sin \left((2L-1)\frac{\pi r}{2T+1}\right)}
       {\sin\left(\frac{\pi r}{2T+1}\right)},    & r \ne 0 ,  \\
  2L-1,                                          & r=0     .  \\
 \end{array}
\right.
\label{eqn:e1}
\end{eqnarray}

When $r\ne 0$,
\begin{eqnarray}
& & \sum_{i=1-L}^{L-1}|i|e^{-j2\pi \frac{ri}{2T+1}} \nonumber \\
&=& \sum_{i=1}^{L-1}ie^{-j2\pi \frac{ri}{2T+1}} +
    \sum_{i=1-L}^{-1}(-i)e^{-j2\pi \frac{ri}{2T+1}} \nonumber \\
&=& j\frac{2T+1}{2\pi}\frac{\partial}{\partial r}
    \sum_{i=1}^{L-1}
    \left(e^{-j2\pi \frac{ri}{2T+1}}
    -e^{j2\pi \frac{ri}{2T+1}}\right) \nonumber \\
&=& j\frac{2T+1}{2\pi}\frac{\partial}{\partial r}
    \frac{\sin \left((L-1)\frac{\pi r}{2T+1}\right)}
         {\sin\left(\frac{\pi r}{2T+1}\right)}
    \left(e^{-j\pi \frac{rL}{2T+1}}-
    e^{j\pi \frac{rL}{2T+1}}\right) \nonumber \\
&=& \frac{L\cos\left((L-1)\frac{2\pi r}{2T+1}\right)-
    (L-1)\cos\left(\frac{2L\pi r}{2T+1}\right)-1}
    {2\sin^2\left(\frac{\pi r}{2T+1}\right)}.
\label{eqn:e2}
\end{eqnarray}

When $r=0$,
\begin{equation}
\sum_{i=1-L}^{L-1}|i|e^{-j2\pi \frac{ri}{2T+1}}=L(L-1).
\label{eqn:e3}
\end{equation}

Substituting (\ref{eqn:Dr})-(\ref{eqn:e3}) into 
(\ref{eqn:Vk2}), we obtain
\begin{eqnarray}
& & V(r) \nonumber \\
&=& \left\{
\begin{array}{ll}
  \frac{2\alpha\sin\left(\frac{\pi r}{2T+1}\right)
  \left(\left(1-U\right)\sin\left(\frac{\pi r}{2T+1}\right)+
  U\sin\left((2L+1)\frac{\pi r}{2T+1}\right)\right)}
{2\sin^2\left(\frac{\pi r}{2T+1}\right)
-U^2\left(1-\cos\left(\frac{2L\pi r}{2T+1}\right)\right)},
   & r \ne 0 ,\\
\frac{\alpha(1+2UL)}{1-U^2L^2},  & r=0 . \\
 \end{array}
\right.
\label{eqn:Vr2}
\end{eqnarray}

Since the inverse discrete Fourier transformation of (\ref{eqn:Vr2})
equals $v\left(n\right)$, we obtain
\begin{equation}
v(n)=\lim_{T\rightarrow \infty}
\frac{1}{2T+1}\sum_{r=-T}^T V(r)e^{j2\pi \frac{rn}{2T+1}} .
\label{eqn:vn}
\end{equation}

Substituting (\ref{eqn:vn}) into (\ref{eqn:sigmasteady}),
we obtain
\begin{eqnarray}
\sigma^2
&=& \lim_{T\rightarrow \infty}\frac{1}{2T+1}\sum_{r=-T}^{T}V\left(r\right)
\sum_{n=1-L}^{L-1}\left(L-|n|\right)e^{j2\pi\frac{rn}{2T+1}} .
\label{eqn:sig2}
\end{eqnarray}

Using (\ref{eqn:e1})-(\ref{eqn:sig2}) 
and rewriting $\frac{r}{2T+1}\rightarrow x$ , 
$\frac{1}{2T+1}\rightarrow dx$, 
we can express $\sigma^2$ as a form using a simple integral like 
(\ref{eqn:sigma2steady}). 
As a result, we can obtain the steady state equations 
in terms of the macroscopic variables of the network
as (\ref{eqn:sigma2steady})-(\ref{eqn:mm}).

\section*{Acknowledgment}
This research was partially supported by the Ministry of Education, 
Culture, Sports, Science and Technology, Japan, 
with Grant-in-Aid for Scientific Research
13780303, 14580438 and 15500151.

\newpage
%

%

\newpage
\section*{Figure Captions}
\begin{enumerate}
\item Structure of delayed network.
\item Relationship between loading rate $\alpha$ and 
overlap $m$ (theory).
\item Relationship between loading rate $\alpha$ and 
overlap $m$ (computer simulation).
\item Relationship between loading rate $\alpha$ and 
overlap $m$. These lines are obtained by solving
steady state equations numerically.
\item Relationship between length of delay $L$ and storage 
capacity $\alpha_C$. This line is obtained by solving the
steady state equations numerically. Storage capacity is $0.195L$,
with a large $L$ limit.
\item Representation of delayed synapses by pruning: 
(a) length of delay is three, and
(b) length of delay is five.
\item Relationship between loading rate $\alpha$ and 
overlap $m$ when synapses are randomly pruned
(theory(t) and computer simulation(s)).
\item Relationship between loading rate $\alpha$ and 
overlap $m$ when synapses are randomly pruned (theory).
\item Relationship between $\log L$ and $\log r$ when synapses 
are randomly pruned.
\item Nonlinear function for systematic pruning.
\item Relationship between loading rate $\alpha$ and 
overlap $m$ when synapses are systematically pruned
(theory(t) and computer simulation(s)).
\item Relationship between loading rate $\alpha$ and 
overlap $m$ when synapses are systematically pruned (theory).
\item Relationship between length of delay $L$ and 
storage capacity $\alpha_C$ when synapses are systematically 
pruned.
\item Relationship between $\log L$ and $\log \sigma^2$ when synapses 
are systematically pruned.
\end{enumerate}

\newpage
\section*{Figures}
\def\graphsize{0.570} 

\begin{figure}[htbp]
\begin{center}
\includegraphics[width=\graphsize\linewidth,keepaspectratio]{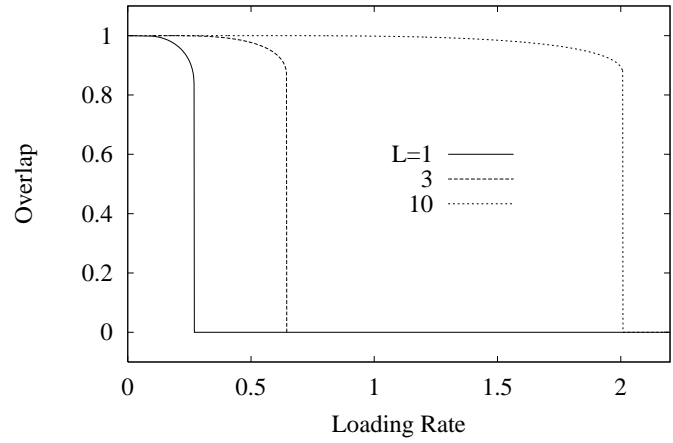}
\caption{Relationship between loading rate $\alpha$ and 
overlap $m$ (theory).}
\label{fig:Lsnocut2}
\end{center}
\end{figure}

\begin{figure}[htbp]
\begin{center}
\includegraphics[width=\graphsize\linewidth,keepaspectratio]{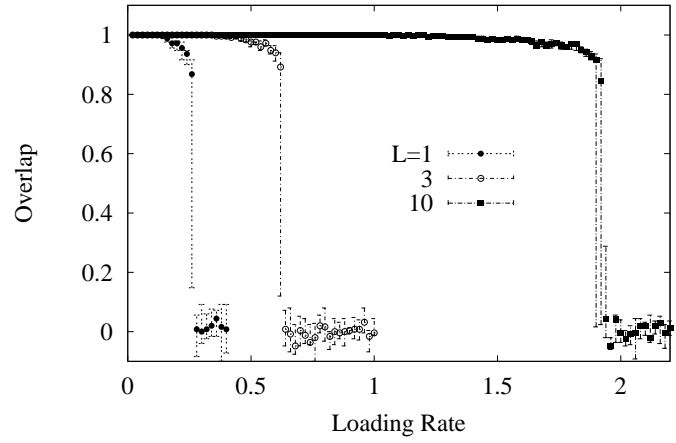}
\caption{Relationship between loading rate $\alpha$ and 
overlap $m$ (computer simulation).}
\label{fig:LdAs}
\end{center}
\end{figure}

\begin{figure}[htbp]
\begin{center}
\includegraphics[width=\graphsize\linewidth,keepaspectratio]{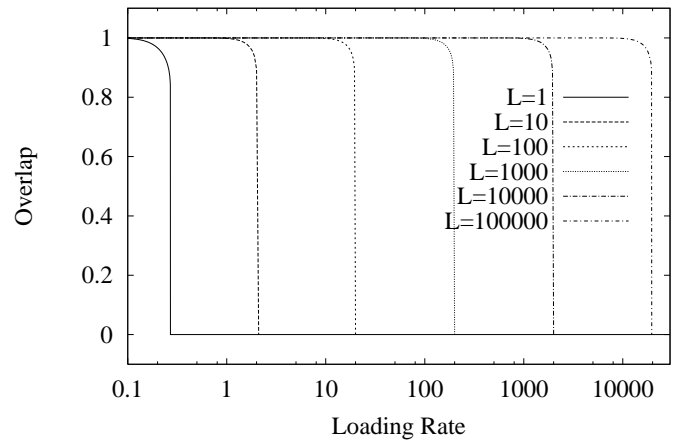}
\caption{Relationship between loading rate $\alpha$ and 
overlap $m$. These lines are obtained by solving
steady state equations numerically.}
\label{fig:Lsnocut}
\end{center}
\end{figure}

\begin{figure}[htbp]
\begin{center}
\includegraphics[width=\graphsize\linewidth,keepaspectratio]{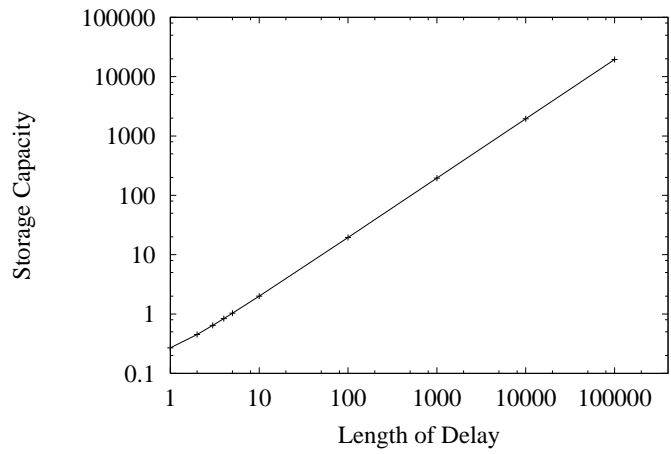}
\caption{Relationship between length of delay $L$ and storage 
capacity $\alpha_C$. This line is obtained by solving the
steady state equations numerically. Storage capacity is $0.195L$,
with a large $L$ limit.}
\label{fig:nocut}
\end{center}
\end{figure}

\begin{figure}[htbp]
\begin{center}
\includegraphics[width=\graphsize\linewidth,keepaspectratio]{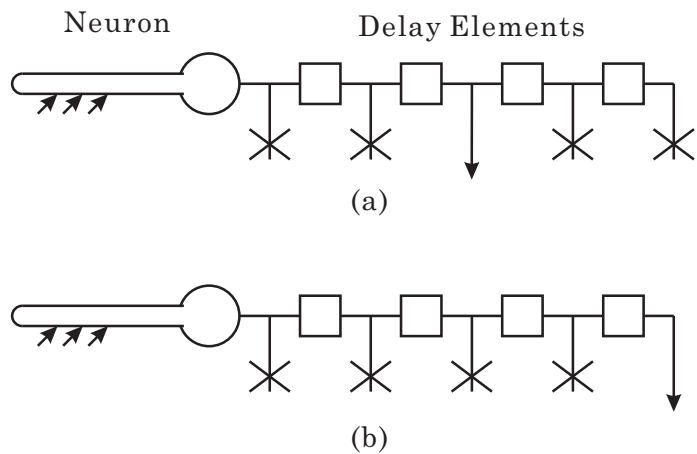}
\caption{Representation of delayed synapses by pruning: 
(a) length of delay is three, and
(b) length of delay is five.}
\label{fig:L5}
\end{center}
\end{figure}

\begin{figure}[htbp]
\begin{center}
\includegraphics[width=\graphsize\linewidth,keepaspectratio]{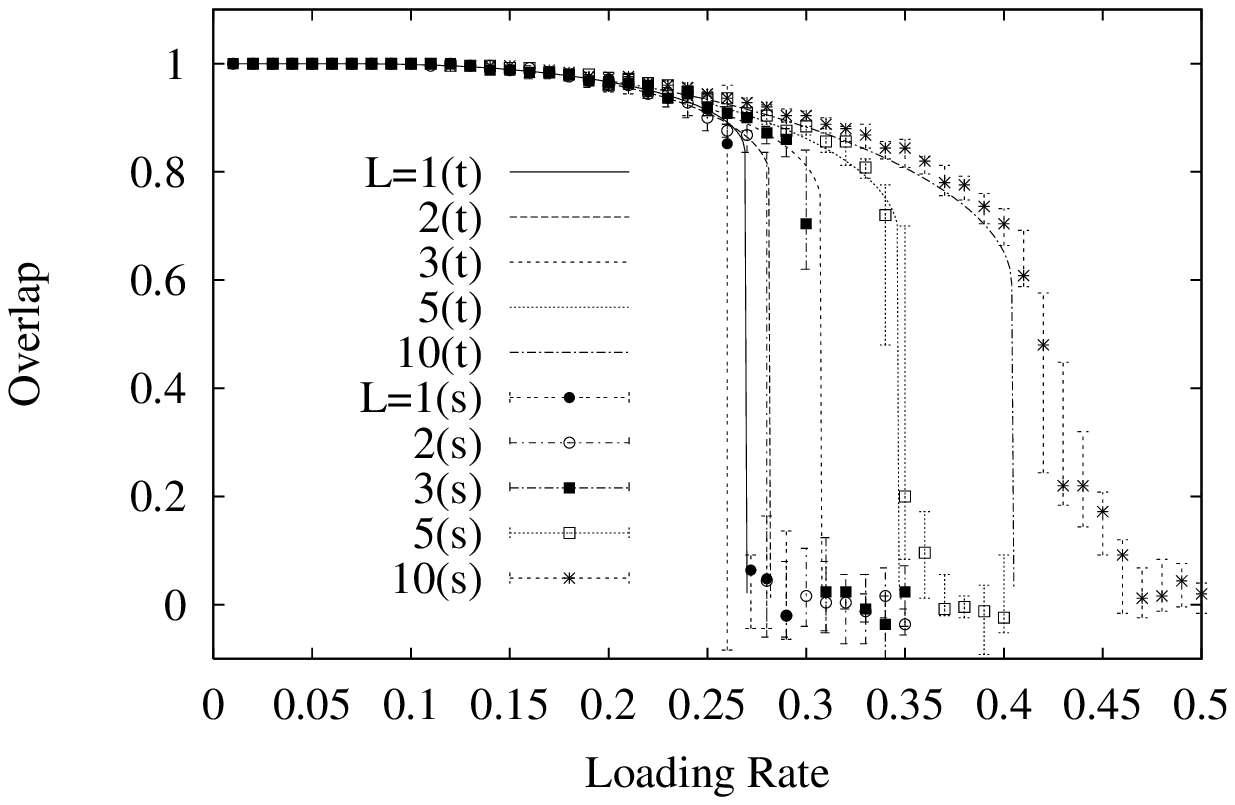}
\caption{Relationship between loading rate $\alpha$ and 
overlap $m$ when synapses are randomly pruned
(theory(t) and computer simulation(s)).}
\label{fig:LArt+s}
\end{center}
\end{figure}

\begin{figure}[htbp]
\begin{center}
\includegraphics[width=\graphsize\linewidth,keepaspectratio]{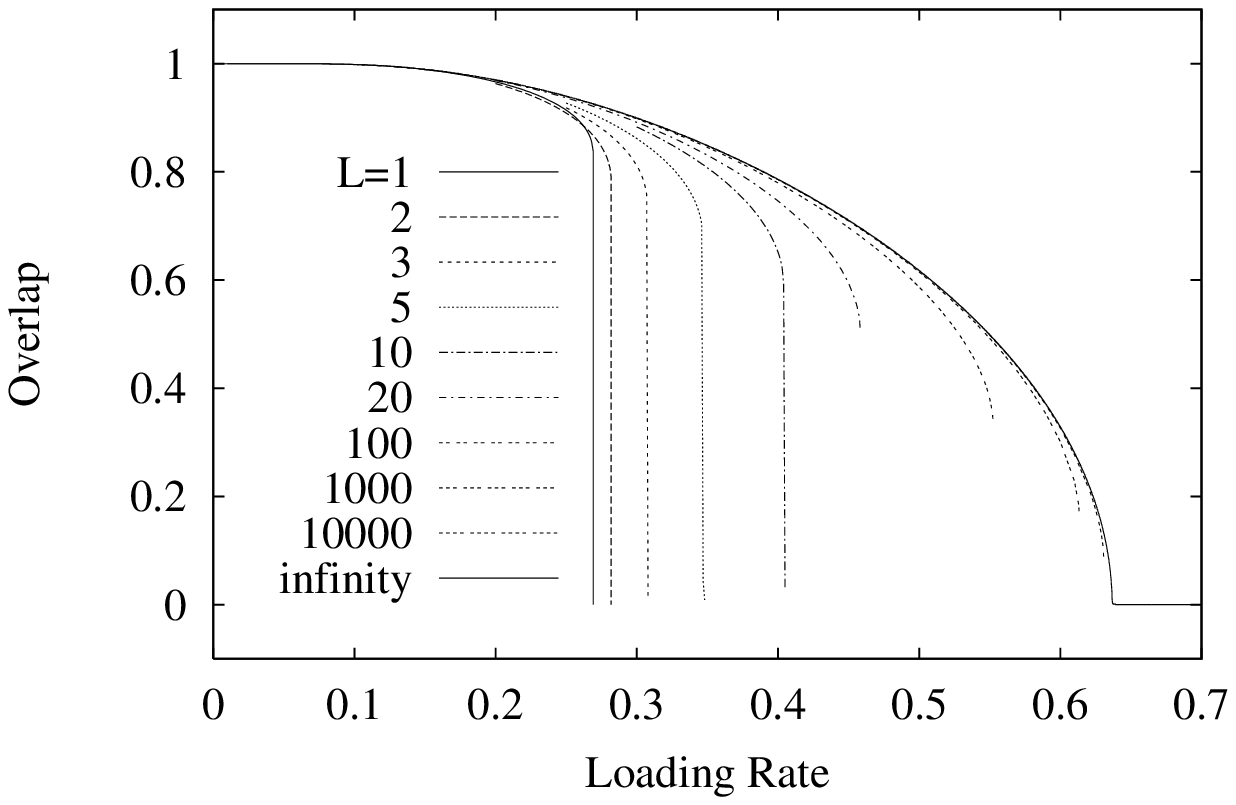}
\caption{Relationship between loading rate $\alpha$ and 
overlap $m$ when synapses are randomly pruned (theory).}
\label{fig:randomt}
\end{center}
\end{figure}

\begin{figure}[htbp]
\begin{center}
\includegraphics[width=\graphsize\linewidth,keepaspectratio]{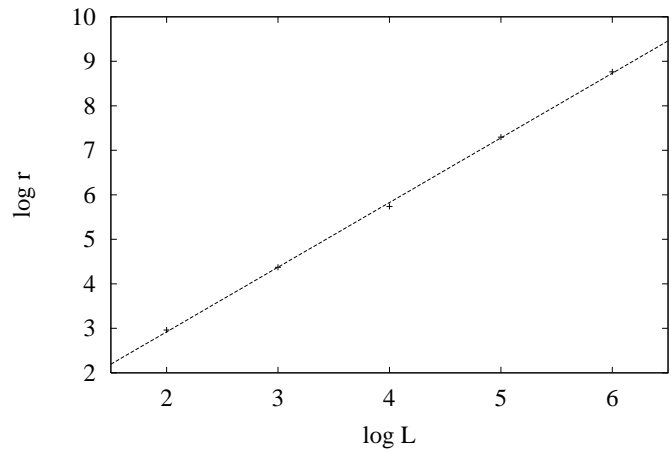}
\caption{Relationship between $\log L$ and $\log r$ when synapses 
are randomly pruned.}
\label{fig:L-term1r}
\end{center}
\end{figure}

\begin{figure}[htbp]
\begin{center}
\includegraphics[width=\graphsize\linewidth,keepaspectratio]{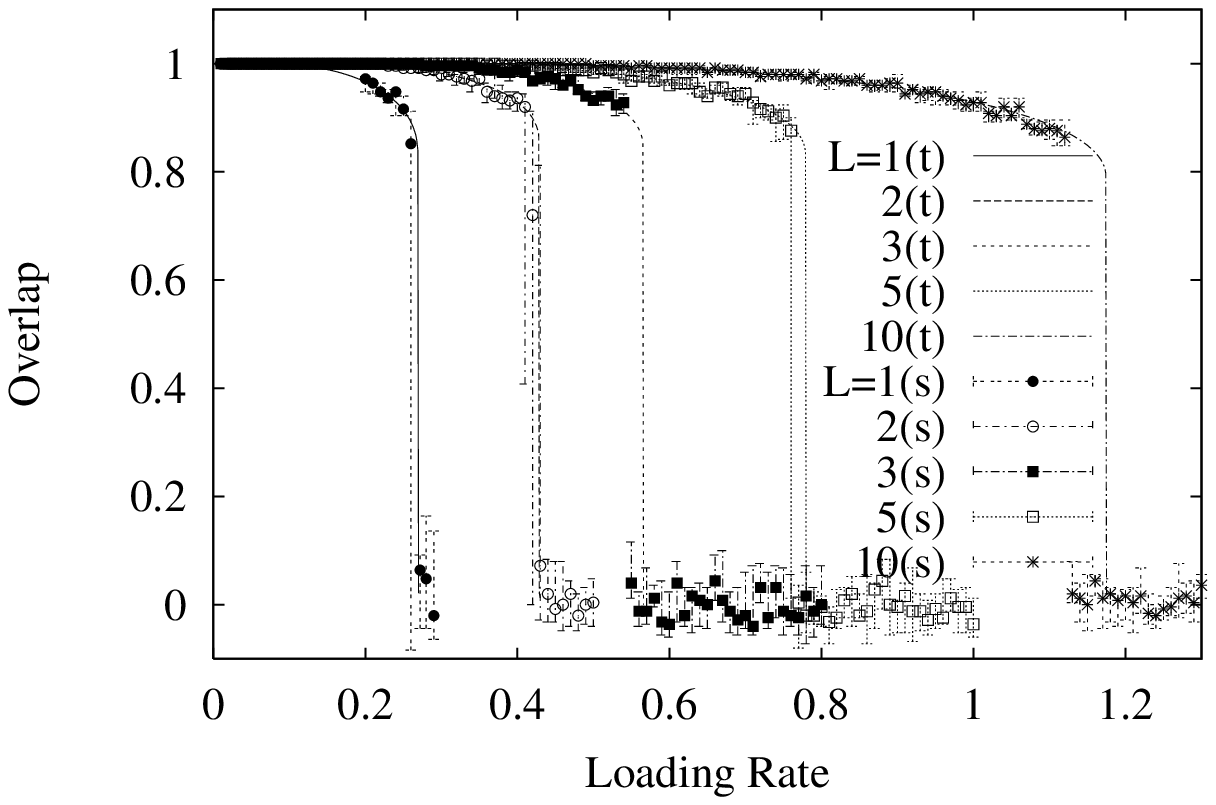}
\caption{Relationship between loading rate $\alpha$ and 
overlap $m$ when synapses are systematically pruned
(theory(t) and computer simulation(s)).}
\label{fig:LAst+s}
\end{center}
\end{figure}

\begin{figure}[htbp]
\begin{center}
\includegraphics[width=\graphsize\linewidth,keepaspectratio]{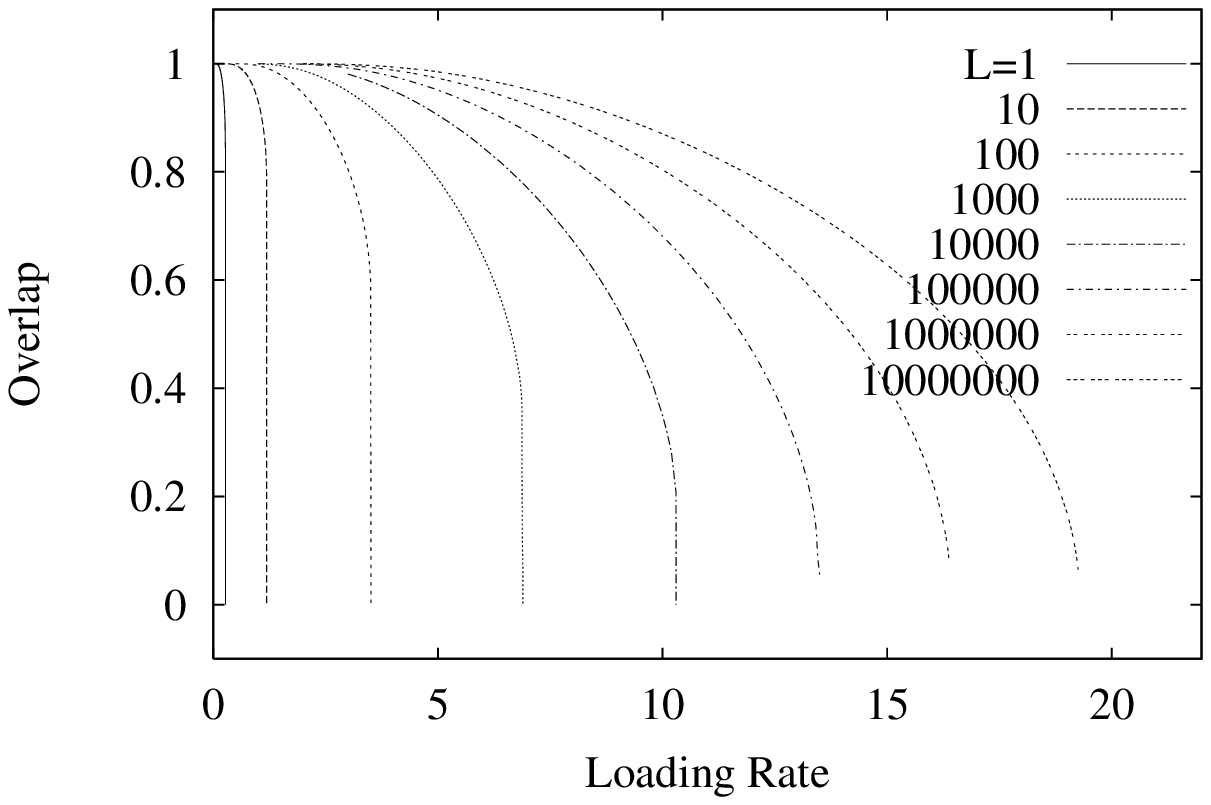}
\caption{Relationship between loading rate $\alpha$ and 
overlap $m$ when synapses are systematically pruned (theory).}
\label{fig:systemt}
\end{center}
\end{figure}

\begin{figure}[htbp]
\begin{center}
\includegraphics[width=\graphsize\linewidth,keepaspectratio]{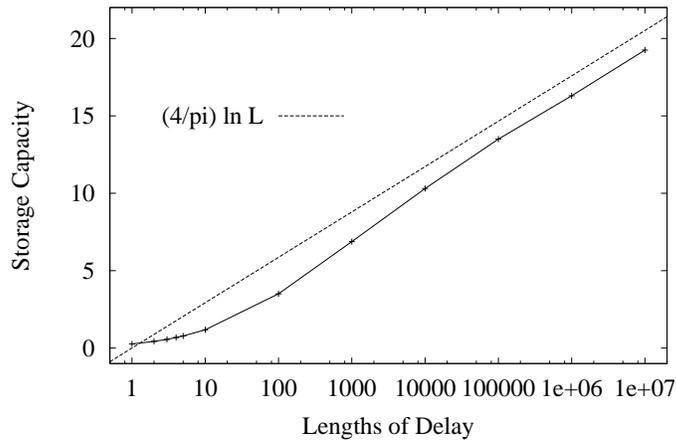}
\caption{Relationship between length of delay $L$ and 
storage capacity $\alpha_C$ when synapses are systematically 
pruned.}
\label{fig:systemcapa}
\end{center}
\end{figure}

\begin{figure}[htbp]
\begin{center}
\includegraphics[width=\graphsize\linewidth,keepaspectratio]{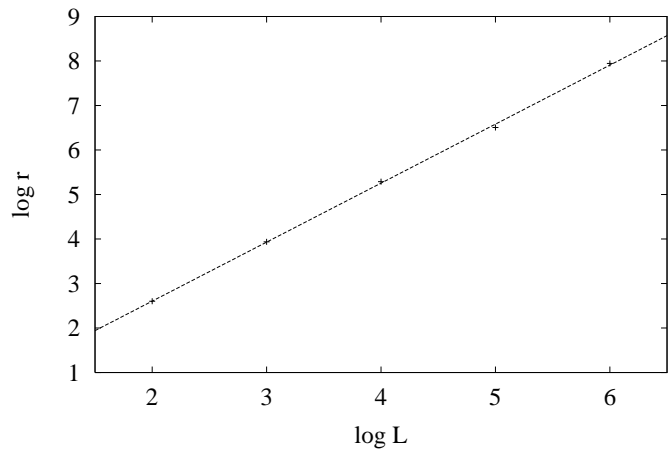}
\caption{Relationship between $\log L$ and $\log \sigma^2$ when synapses 
are systematically pruned.}
\label{fig:L-term1s}
\end{center}
\end{figure}



\end{document}